\documentclass[conference]{IEEEtran}
\def\ps@headings{%
\def\@oddhead{\mbox{}\scriptsize\rightmark \hfil \thepage}%
\def\@evenhead{\scriptsize\thepage \hfil \leftmark\mbox{}}%
\def\@oddfoot{}%
\def\@evenfoot{}}
\makeatother
\usepackage{booktabs} 
\usepackage{url}
\usepackage{svg}
\usepackage{etoolbox}
\usepackage{cuted}
\usepackage{lipsum}
\usepackage{amssymb}
\usepackage{array}
\usepackage{amsmath}
\usepackage{tikz}
\usepackage{color}
\usepackage{dirtytalk}
\usepackage{algpseudocode}
\usepackage{graphicx}
\usepackage{caption}
\usepackage{mdframed}
\usepackage{tcolorbox}
\usepackage{comment}
\usepackage{mathtools}

\usepackage{pifont}
\usepackage{booktabs,makecell}
    
\usepackage{siunitx}

\usepackage[utf8x]{inputenc}
\usepackage{amsmath,amsthm}

\usepackage{multirow}
\usepackage[ruled,vlined]{algorithm2e}
\usepackage{fancyvrb}

\theoremstyle{definition}

\theoremstyle{plain}

\ifCLASSOPTIONcompsoc
  \usepackage[nocompress]{cite}
\else
  \usepackage{cite}
\fi
\ifCLASSINFOpdf
\else
\fi
\begin{document}
%
\title{A Location-based and Hierarchical Framework for Fast Consensus in Blockchain Networks}

\author{\IEEEauthorblockN{Hao Guo\IEEEauthorrefmark{1}~~~~Wanxin Li\IEEEauthorrefmark{2}~~~~Mark Nejad\IEEEauthorrefmark{2}}
\IEEEauthorblockA{\IEEEauthorrefmark{1}School of Software, Northwestern Polytechnical University, Taicang Campus, China. \\ \IEEEauthorrefmark{2}
Department of Civil and Environmental Engineering, University of Delaware, U.S.A.\\
haoguo@nwpu.edu.cn \& {\{wanxinli,nejad\}@udel.edu}}


}
\maketitle
\pagestyle{headings} 

\begin{abstract}


Blockchain-based IoT systems can manage IoT devices and achieve a high level of data integrity, security, and provenance. However, incorporating the existing consensus protocols in many IoT systems limits scalability and leads to high computational cost and network latency.
We propose a hierarchical and location-aware consensus protocol for IoT-blockchain applications inspired by the original Raft protocol to address these limitations.  
 The proposed consensus protocol generates the consensus candidate groups based on nodes' individual reputation and distance information to elect the leader in each sub-layer blockchain and uses our threshold signature scheme to reach global consensus. 
 Experimental results show that the proposed consensus protocol is scalable for large IoT applications and significantly reduces the communication cost, network latency, and agreement time by more than 50\% compared with the Raft protocol for consensus processing.
\end{abstract}
\IEEEpeerreviewmaketitle

\section{Introduction}
\label{sec:introduction}

Modern IoT networks are often large-scale, dynamically located, and globally distributed. By 2025, IoT devices such as smart home appliances, smartphones, and other types of smart sensors will increase to more than 75 billion~\cite{iotnumber}. Many IoT networks require massive data communication and need to manage unreliable and failure messages, among others, automatically. Consensus protocols promise to achieve overall system reliability in the presence of inconsistent and failure messages by coordinating processes to reach agreements.  State machine replication (SMR) is a fundamental method for system availability and fault tolerance in the distributed systems. For example, Autopilot~\cite{isard2007autopilot} builds fault-tolerant replicas in Microsoft's data centers worldwide using consensus protocols. Google File System utilizes the Chubby~\cite{burrows2006chubby} lock service to reach the consensus for the replication of different files.  However, there are significant challenges in the current IoT applications, including data integrity, resource-intensive consensus mechanisms, high latency, and limited scalability~\cite{putra2021trust}.

 
 Blockchain is a distributed ledger that records transactions among multiple participants in a verifiable manner. Blockchain can reduce the costs involved in verifying transactions as a distributed ledger by removing the need for a trusted third-party operating as a centralized authority.  Since the introduction of Bitcoin~\cite{nakamoto2008bitcoin}, blockchain applications have expanded beyond cryptocurrencies and financial-related fields. The smart contract's invention \cite{ethofficial, wood2014ethereum} leads to developing more varied applications, such as blockchain-based intelligent transportation systems (e.g., \cite{li2020privacy, li2020blockchain, guo2020proof, li2021ICBC}) and smart health (e.g., \cite{guo2020icbc, guo2019access}). However, blockchains with a complex application layer and smart contracts can incur significant computation for transaction execution.

Many IoT-blockchain applications, such as for managing the electric power grid \cite{lu2021edge}, can benefit from a hierarchical architecture to reduce the consensus process and data communication time. A hierarchical multi-layer blockchain network can communicate within its sub-layers and achieve the consensus in a more efficient way~\cite{li2020scalable}. In this paper, we propose a novel consensus protocol for blockchain-based IoT applications, which is inspired by the original Raft protocol~\cite{ongaro2014search}. 
By incorporating the geographic information of the IoT devices,  our propose consensus protocol boosts the blockchain performance and makes the system more dynamic and  immune to the Sybil attacks~\cite{douceur2002sybil}. 

In each sub-layer blockchain network, our consensus protocol engages a few local nodes with a low consensus latency, making the blockchain-based IoT system more efficient. Our scheme generates candidate groups based on nodes' reputation and distance information to elect the leaders. Our design solution partitions the blockchain network into a hierarchical structure of sub-blockchains with a threshold signature scheme to reach the global consensus to achieve higher system scalability.

%
%
%
%




Specifically, we design the novel consensus protocol, which constructs sub-layers  {\it local consensus} based on IoT devices' reputation and geographic information to elect the leader, and construct a hierarchical architecture by utilizing a threshold signature scheme to reach {\it global consensus} among the multiple layers. 
We evaluate the new consensus protocol's performance 
and compare it with the classical Raft protocol. 
 Moreover, we experiment with a threshold signature scheme for both signing and verification time 
 using the Hyperledger Ursa cryptography library \cite{hyperledgerursa}. Compared to the original Raft protocol, our proposed scheme reach consensus significantly 
 faster with lower network overhead. 

The rest of the paper is organized as follows. We discuss the related work in Section II. In Section III, we describe the system architecture. In addition, we present the threshold signature scheme and location-based hierarchical raft protocol. In Section IV, we conducted experiments and evaluations based on our proposed new consensus protocol.  In Section V, we conclude the paper and point out future research directions.

\section{Related Work}
Yu et al.~\cite{yu2020layerchain} proposed a hierarchical edge-cloud blockchain architecture named LayerChain. They described a layered structure to save the blockchain transaction data in multiple distributed clouds and edge nodes. Moreover, to mitigate lengthy delays during block propagation, they developed a tree-based clustering algorithm where blocks are propagated via different clusters with a compressed tree depth.  Chuang et al.~\cite{chuang2020hierarchical} proposed a hierarchical blockchain-based data service platform in MEC environments. This system provided an adaptive PoW consensus scheme that dynamically changed the hash puzzle's difficulty and enhanced resource-constrained IoT devices. 
Yang et al.~\cite{yang2019blockchain} proposed a hierarchical trust networking architecture to implement JointCloud (HTJC). By developing the credit bonus-penalty strategy (CBPS), HTJC can address the trust issue and provide participants with a secure and trusted trade environment. 

Lao et al.~\cite{lao2020g} proposed G-PBFT (Geographic-PBFT), a location-based and scalable consensus mechanism for IoT-blockchain applications. In their design, G-PBFT utilized the era switch mechanism to maintain the dynamics in the IoT devices. The experiment results showed that G-PBFT reduced the network overhead and consensus time significantly. 
Li et al.~\cite{li2020scalable} proposed a scalable multi-layer PBFT consensus protocol for blockchain. The proposed double-layer PBFT scheme reduces the communication complexity significantly. They also analyzed the security threshold based on the faulty probability determined and the faulty number determined models. An et al.~\cite{an2019tcns} proposed a decentralized privacy-preserving model based on the twice verifications process and consensuses of the blockchain system. They introduced a twice consensus mechanism, ensuring that data can be traced and prevented from being impersonated and denied.

\section{System Architecture}
\begin{figure}[t]
\centering
\includegraphics[width=0.488\textwidth]{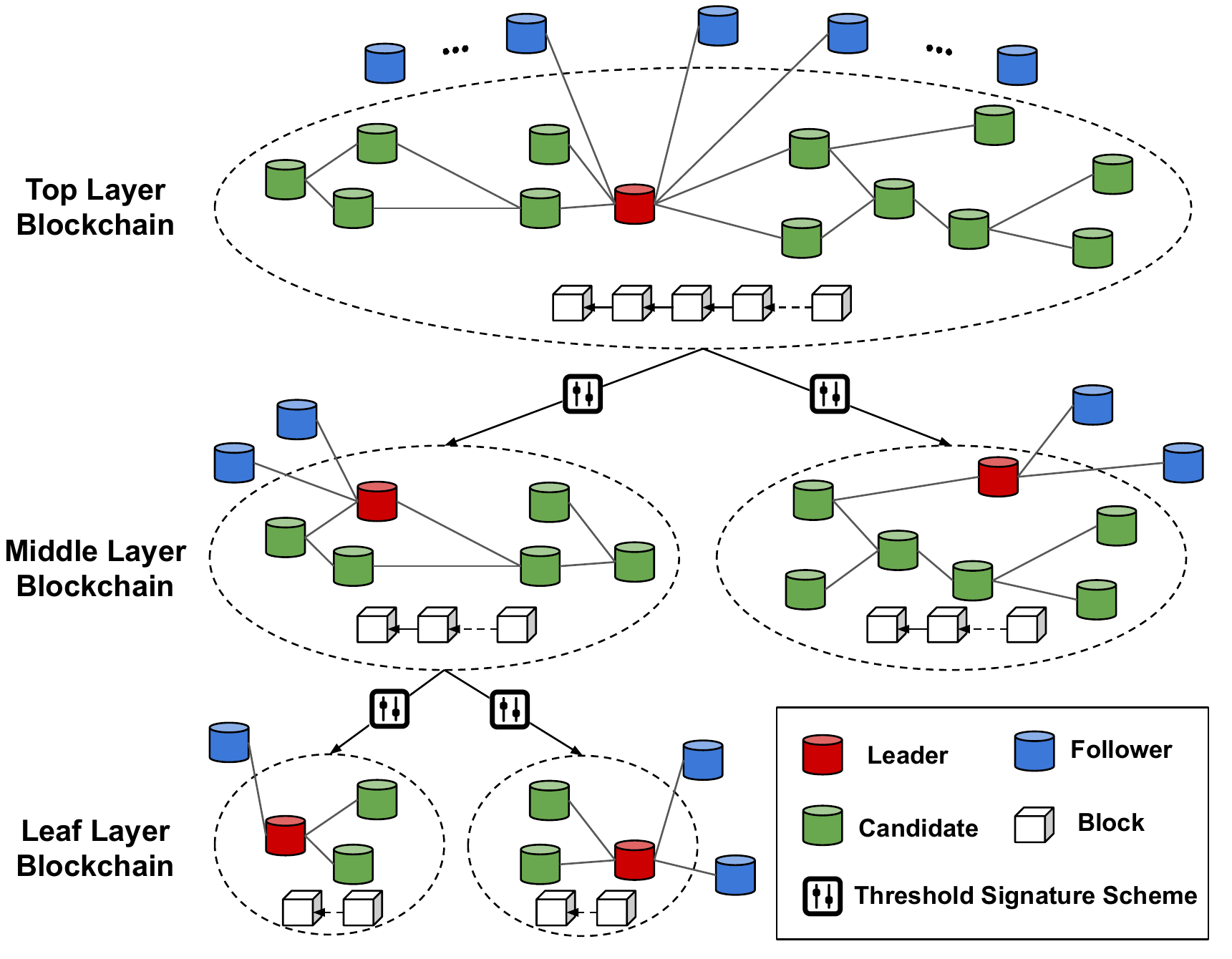}
\caption{Location-based Hierarchical Blockchain Architecture for IoT Applications.}
\label{fig:hierarchical}
\end{figure}

The underlying location-centric characteristics inherent in many IoT applications necessitates location-aware design solutions. This section describes our proposed system 
architecture for location-aware consensus in IoT networks.  
In our approach, 
the global network is divided into sub-blockchains based on regional information. These sub-blockchains are connected in a hierarchical structure forming the more extensive global blockchain network. The system provides a multi-chain and multi-level structure, as shown in Fig.~\ref{fig:hierarchical}. 
We first define the following entities that take part in the proposed architecture.

\begin{itemize}
\item 
Blockchain: Blockchain serves as the coordinator for IoT devices and manages data sharing and access activities. It forms a dynamic hierarchical structure based on the IoT device's geographic information, including top, middle, and leaf layers.

\item IoT Nodes: The IoT nodes participate in the consensus process. There are three types of IoT nodes: {\it Follower}, {\it Candidate}, and {\it Leader} nodes. They can switch the role seamlessly between these statuses. All {\it Candidate} nodes together form the consensus nodes group, which elect the {\it Leader} node.

\item Client: A client node only requests new transactions to append data to the ledger, and they do not participate in the consensus procedure. For example, a healthcare system's smart devices can host client nodes to request electronic health record updates. 

\item Threshold Signature Scheme: The threshold signature scheme is proposed to achieve consensus among hierarchical blockchain layers in the architecture. We will explain the detailed construction in the following subsections.


\end{itemize}

\begin{figure}[t]
\centering
\includegraphics[width=0.488\textwidth]{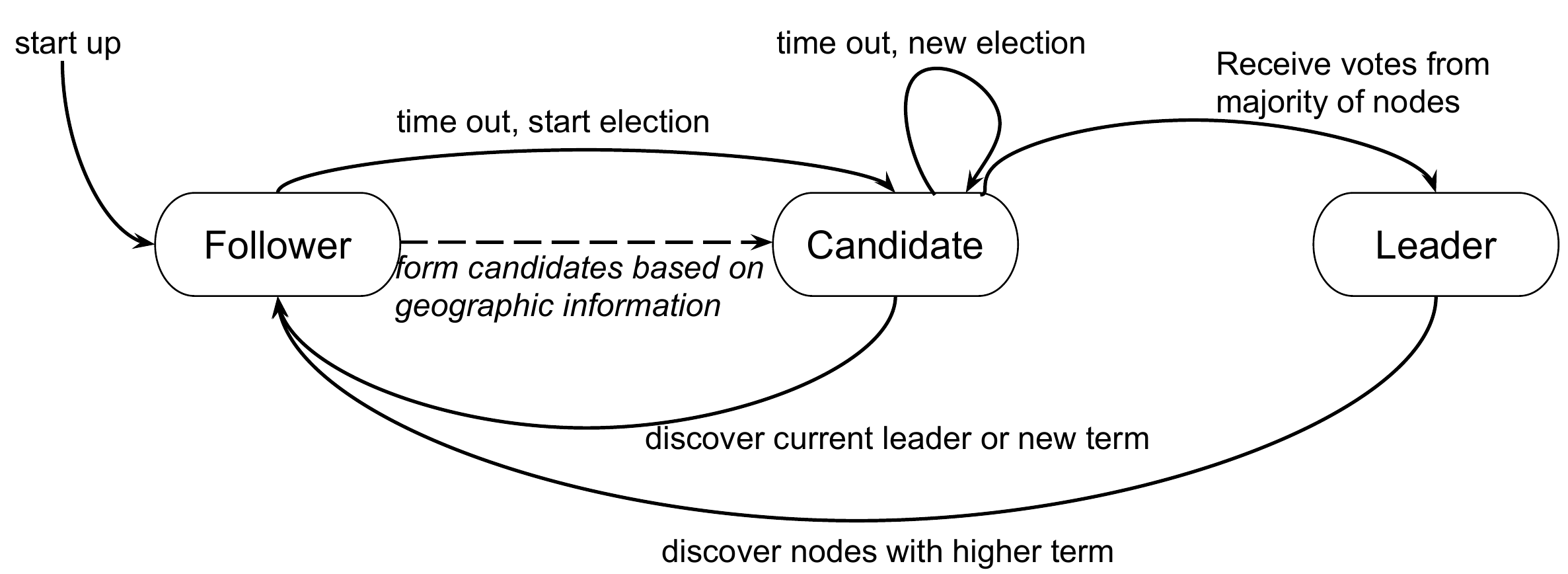}
\caption{Location-based Hierarchical Raft Protocol.}
\label{fig:lhraft}
\end{figure}

As shown in Fig.~\ref{fig:lhraft}, the proposed location-based hierarchical raft protocol has three participant entities: {\it Follower}, {\it Candidate}, and {\it Leader}. These interconvertible nodes can change their status to other roles. The {\it Candidate} and {\it Leader} nodes participate in the consensus process. The {\it Leader} nodes maintain the integrity and confidentiality of the blockchain system and broadcast the newly generated transactions to the {\it Follower} nodes. By contrast, the {\it Follower} nodes will only start new election processes and form {\it Candidate} nodes groups based on their geographic information. Transactions are determined among {\it Candidate} and {\it Leader} nodes to reduce communication overhead. If any message is failed, a node will resend the message again after the timeout period. The role of a node in our proposed scheme is not fixed; a {\it Follower} node can become the {\it Candidate} node, and {\it Candidate} node can become the {\it Leader} node. On the other hand, if the location of a {\it Leader} node has been changed or it conducts malicious action, it can be detected by the voting process by the {\it Candidate} nodes. 

The proposed architecture is designed in a way that is not affected by the size of the IoT network. Rather than all nodes participating in the consensus procedure, nodes execute {\it local consensus} within each sub-layer blockchain. Each sub-layer blockchain leader engages in the hierarchical consensus by utilizing the threshold signature scheme to reach {\it global consensus}.

In the remainder of this section, we first  present the threshold signature scheme to reach the consensus between multiple blockchain layers in the hierarchical architecture. Next, we describe the location-aware hierarchical raft protocol with detailed constructions. 

\subsection{Threshold Signature Scheme}
We proposed the threshold signature scheme for trust-based hierarchical coordination and operation between two blockchain layers. For example, assume that a lower layer blockchain network contains $n$ consensus nodes and the threshold is set as $t$-out-of-$n$ between the lower layer and the upper layer blockchain network. The upper layer blockchain, acting as the verifier network, will trust this lower layer blockchain only if at least $t$ consensus nodes' signatures are verified as legitimate. In addition, utilizing bilinear pairing-based cryptography, the signatures can be verified without disclosing any sensitive information.

\begin{algorithm}[ht]
\label{alg: zkp-keygen}
\SetAlgoLined
\LinesNumbered
\SetKwInOut{Input}{Input}
\SetKwInOut{Output}{Output}
\Input{For each consensus node $i$}
\Output{Verifier key $v_i$}
The permission issuer selects a random $a_i \in \mathbb{Z}_p$\ for consensus node $i$ \;
The permission issuer computes the verifier key as $v_i = g^{a_i} \in G$ \;
The permission issuer returns $v_i$ \;
\caption{Key Generation}
\end{algorithm}

\begin{algorithm}[ht]
\label{alg: zkp-proofgen}
\SetAlgoLined
\LinesNumbered
\SetKwInOut{Input}{Input}
\SetKwInOut{Output}{Output}
\Input{Each consensus node's MAC address $m_i$}
\Output{One-time signature $\delta_i$}
The system computes a hash digest $h_i$ based on MAC address $m_i$ via \cite{rachmawati2018comparative}, as $h_i = H(m_i)$ \;
The system generates the one-time signature $\delta_i = {h_i}^{a_i} \in G$ \;
The system sends $\delta_i$ to upper layer network \;
\caption{Threshold Signature Generation}
\end{algorithm}

\begin{algorithm}[ht]
\label{alg: zkp-proofverify}
\SetAlgoLined
\LinesNumbered
\SetKwInOut{Input}{Input}
\SetKwInOut{Output}{Output}
\Input{One-time signature $\delta_i$, hashed MAC address $h_i$, verifier key $v_i$}
\Output{Identity verification result $r$}
k = 0;

\For{
each one-time signature $\delta_i$
}{\eIf{$e(\delta_i, g) == e(h_i, v_i)$}{$r_i = True$ \; $k = k + 1$\;}{$r_i = False$ \;}}
the system checks
\eIf{$k \geq t$}{$r = True$ \;}{$r = False$ \;}
The system returns $r$ \;
\caption{Threshold Signature Verification}
\end{algorithm}

\vspace{3mm}
\noindent

We describe the procedure of the proposed threshold signature scheme in Algorithms 1, 2, and 3. The algorithm contains three main functions: Algorithm 1 shows the key generation function by the system administrator; Algorithm 2 describes the threshold signature generation by lower layer blockchain nodes; Algorithm 3 presents threshold signature verification process by upper layer blockchain nodes, and a threshold $(t,~n)$ is considered in verifying the lower layer blockchain network. The detailed construction of the proposed threshold signature scheme is shown as below:

 \textit{Initial Setup}:  The system has the bilinear pairing function $e$: $\mathbb{G}_1\times \mathbb{G}_2 \rightarrow \mathbb{G}_T$, the secure hash function $H:M \rightarrow \mathbb{G}_1$, and the $(\mathbb{G}_1, \mathbb{G}_2, \mathbb{G}_T, e, g_1, g_2, p, h)$ represents the public parameters.

\textit{Key Generation}: A trusted authority generates signing-verifying key pairs for all consensus nodes in this step. The key generation function selects a random integer $a_i$ as the signing key and computes $g^{a_i}$ as the verifying key for the consensus node $i$.

\textit{Signing}: Each consensus node $i$ computes its hashed identity information $m_i$ as $h_i = H (m_i)$, where $H$ is a hash function such as SHA256 algorithm \cite{rachmawati2018comparative}. Then, this consensus node generates the one-time signature $\delta_i = {h_i}^{a_i}$ and sends it to the upper layer blockchain network.

\textit{Verifying}: Given the one-time signature $\delta_i$ and the verifying key $v_i$, the upper layer blockchain network can verify that $e(\delta_i, g) = e(h_i, v_i)$. This holds because $e({h_i}^{a_i}, g) = e(h_i, g^{a_i}) = e(h_i, g)^{a_i}$ due to the {\it Bilinearity}. Based on the threshold condition, the validity of the lower layer blockchain network $r$ is verified only if at least $t$-out-of-$n$ consensus nodes' one-time signatures $r_i~(1 \leq i \leq n)$ are verified, we write $(r_1, r_2, ..., r_n) {\xrightarrow{(t,n)} r}$.

$\it (t, n)$ threshold signature scheme has been applied based on the BLS signature scheme, where $1\leq t \leq n$.
For instance, If consensus nodes generated three different signatures related to the identity and location information. The upper layer verifier node, which verifies the generated signature $\delta_i$ for consensus node $\it i$, will follow the threshold of {\em 2 out of 3} to authenticate the identity and location information.  Also, the verifier node can apply the threshold signature scheme dynamically, such as the {\em 1 out of 3} rule, to provide more flexibility on the trust between the upper and lower layer blockchain network. 

Compare to the original {\it Raft} protocol, our proposed new consensus algorithm could tolerate the crash and byzantine fault. Classical {\it Raft} protocol requires that all participant nodes are honest and conduct truthful action. By applying the {\it t-of-n} threshold scheme, the verifier node could check and verify the generated signatures independently with fault-tolerance property and protecting the privacy of candidate nodes.





\subsection{Location-based Hierarchical Raft Protocol}
We describe the process of the location-based hierarchical raft protocol in Algorithm \ref{algorithm1}. The algorithm contains six primary phases: lines 2-7 shows the startup of election by a follower node; lines 8-11 describes the nearby node $n_i$ ($\forall$ $n_i$; $n_i$ $\in~N$) sends its candidate group formation (CGF) score (including both reputation and geographical information) to the follower node; lines 12-15 presents the follower node $F$ sorts the $CGF_i$ score~$\forall i$ $~\in~\mathcal{N}$, and chooses the top M nodes to form the candidate group $C$ and broadcasts the candidate group $C$ information, lines 16-18 indicates the confirmation information of candidate group, lines 19-22 shows the confirmation of the elected leader, and finally lines 23-26 presents the current leader switches its role to the follower.

\begin{algorithm}[ht]
\begin{algorithmic}[1]
\State {\bf OUTPUT:}  The $Leader$ of consensus Protocol  
\State {\textbf{START UP}} {\em Follower node $F$ begins the Election.} 
\State {$F$ sends $FormGroup$ request to all nearby nodes $N$;}
 \State $F$ waits for messages in a time period T;
 \State {$F$ gets $CGF$ scores from all nearby nodes $N$;}

  
  \Comment{include reputation and geographical score for each node $n_i~\in~N$}

  \If {no $answer$ from $N$ within time $T$} 
  \State  $F$ restarts  $Election$ procedure;
  \State {\textbf{END START UP}} 
 \State {\textbf{UPON EVENT}} {\em Nearby nodes $N$ receive the FormGroup message:}
 \State  $N$ calculate  $CGF_i$ score for each $n_i$;
 \State each $n_i$ sends $CGF_i$ score to $F$;
 \State {\textbf{END UPON EVENT}}
 \State {\textbf{UPON EVENT}} {\em Follower node $F$ receives the $CGF$ score from all nearby nodes $N$:}
 \State $F$ sorts all $CGF_i$ scores 
 and chooses the top $M$ nodes;
 \State $F$ broadcasts candidates group $C$ to all followers $F$;
  
\State {\textbf{END UPON EVENT}}
\State {\textbf{UPON EVENT}} {\em Nearby nodes $N$ receive the Follower message:}
 \State $N$ accept the nodes $M$ in candidate group $C$;
   \State {\textbf{END UPON EVENT}}
\State {\textbf{UPON EVENT}} {\em Candidates $C$ reach timeout and starts new $Election$:}

   \If{$C_i$ receives $majority$ votes}
  \State $C_i$ becomes the $Leader$ and broadcasts the $Leader$ message to $Followers$ and $Candidates$;
  
  \Else
  \State $C_i$ waits for $Leader$ message from other candidates $C$;
     \State {\textbf{END UPON EVENT}}
     
    \State {\textbf{UPON EVENT}} {\em Leader $L_i$ discovers nodes with higher term:}
    \State $L_i$ accepts other node $L_j$ as the new $Leader$;
    \State $L_i$ switches its role to $Follower$;
 \State {\textbf{END UPON EVENT}}
\caption{Proposed Consensus Protocol}
\label{algorithm1}
\end{algorithmic}
\end{algorithm}



Each IoT device must periodically submit its location, reputation, and timestamp information in the new consensus protocol. We utilize the Crypto-Spatial Coordinates (CSC) mechanism to connect geographical information of IoT devices~\cite{lao2020g}. With CSC, IoT devices could have access to their historical location information. After the qualified IoT node is elected as the {\it Leader} node, it will start to validate and generate a new block and manage blockchain new transaction based on our proposed consensus protocol. If there is a missing block or forking issue caused by the {\it Leader} node, {\it Leader} node will be removed from its {\it Leader} status.



To become the qualified {\it candidate} nodes, the {\it follower} nodes need to satisfy the geographic location requirements. Therefore, the new consensus protocol will check the geographic information of {\it follower} nodes periodically. It will determine if the {\it follower} nodes are within a particular geographic area and whether the node changes its location over a while. If a node's geographic location information has been changed significantly over the past period {\it t}, it will be removed from the {\it candidate} group.

\section{Experiments and Evaluations}
We developed the IoT blockchain system prototype with the new consensus protocol and compared our proposed protocol with the classical Raft protocol. 

\subsection{Communication Cost}

\begin{figure}[t]
\centering
\includegraphics[width=0.423\textwidth]{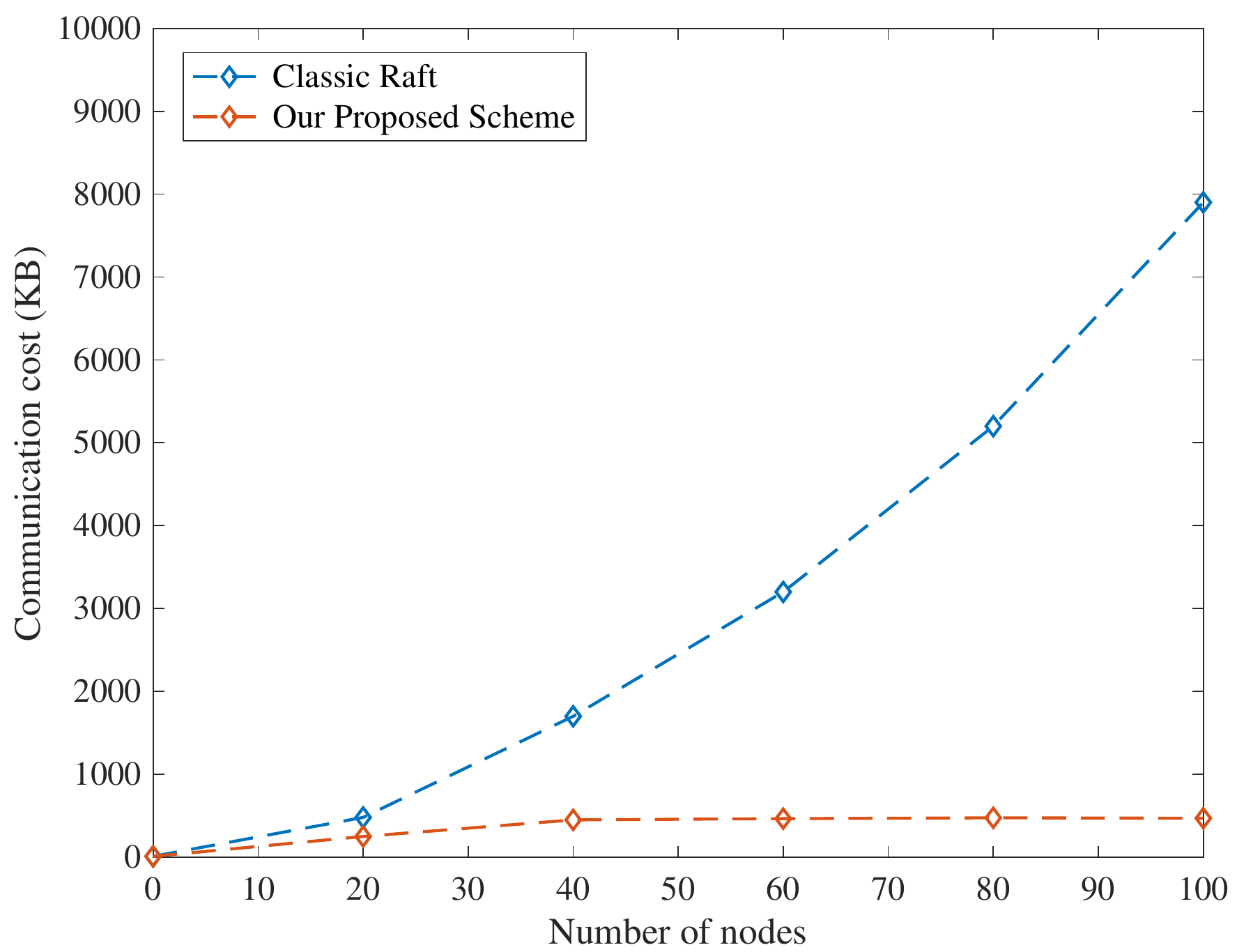}
\caption{Our proposed scheme communication costs vs. classical Raft communication costs.}
\label{fig:commcost}
\end{figure}

Our proposed consensus protocol could reduce the communication cost significantly when the number of IoT devices is large. We set the sub-layer blockchain's {\it Candidate} nodes to be 20 at maximum and evaluate the communication cost for a single blockchain transaction. As shown in Fig.~\ref{fig:commcost}, communication cost in the classical Raft protocol keeps increasing when the number of nodes increases. Moreover, the larger is the number of nodes participating in the system, the more significant is the increase in communication cost. However, for our proposed consensus protocol, the communication cost reaches the upper bound of about 470kb since the sub-layer blockchain has the maximum capability for {\it Candidate} nodes.

Similar to the network latency analysis, the classical Raft protocol cannot work well when candidate nodes are greater than 100. The blue dashed line representing the classical Raft protocol has the 8000kb communication cost when nodes are 100. In contrast, our proposed new consensus protocol can reduce the communication cost to 5.07$\%$ as observed from Fig.~\ref{fig:commcost}.

\subsection{Network Latency}
\begin{figure}[t]
\centering
\includegraphics[width=0.42\textwidth]{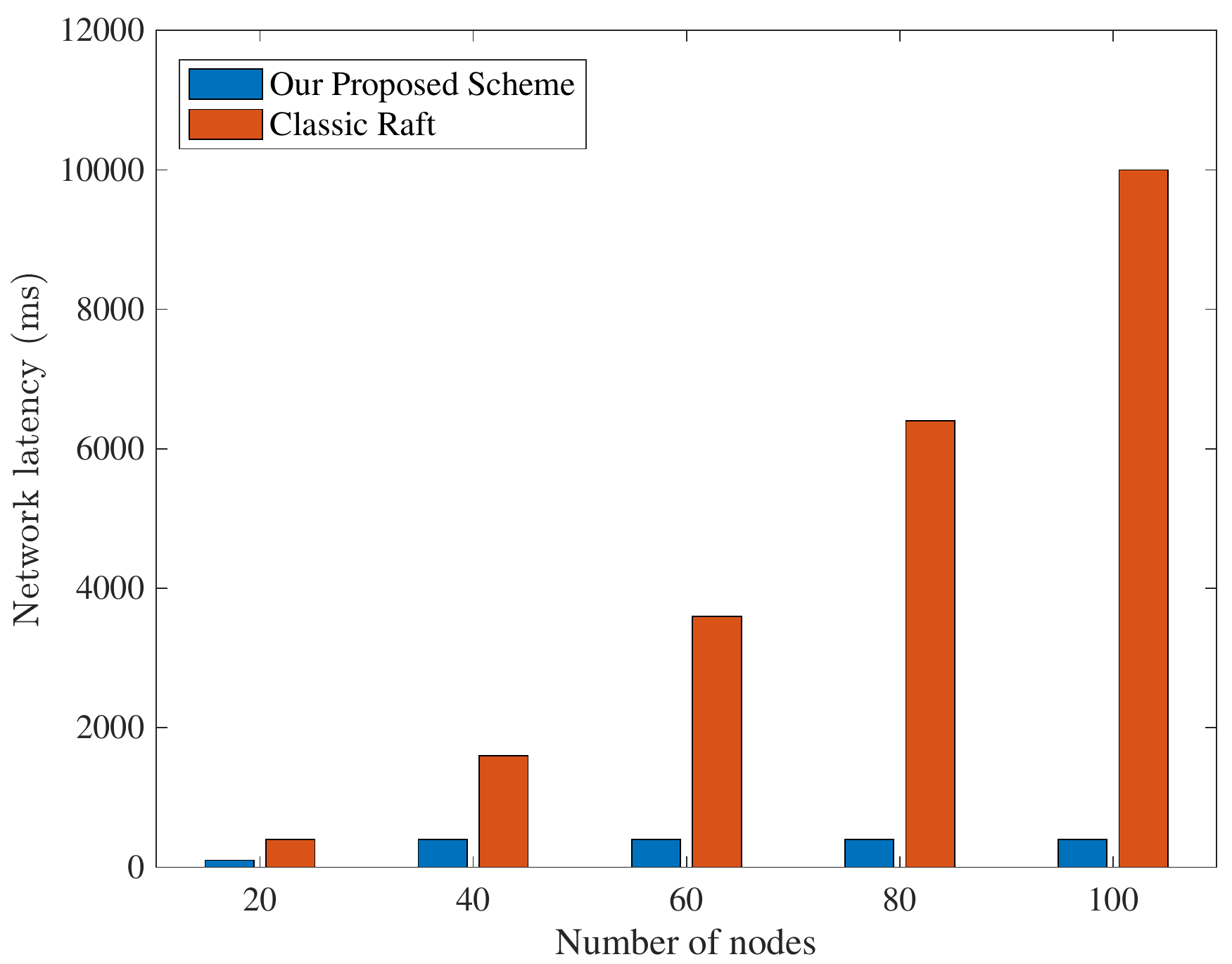}
\caption{Network latency in classical Raft vs. our propose scheme.}
\label{fig:raftlatency}
\end{figure}

In this subsection, we compare the network latency results of our consensus protocol and classical Raft protocol. We pick one {\it Follower} node randomly per sub-layer blockchain.  As shown in Fig.~\ref{fig:raftlatency}, our consensus protocol showed significant latency improvements over the classical Raft by increasing the number of nodes (e.g., {\it 2.1x} network latency decrease for 20 nodes). Note that the maximum number of candidates group is 20 here. Compared to the classical Raft, our proposed approach selects a smaller number of {\it Candidate} nodes based on the geographic information. When nodes increase from 20 to 100, network latency increases exponentially in the classical Raft protocol.



By contrast, the new consensus protocol performs a better performance in terms of the network latency result. All qualified nodes can join the consensus committee when the number of nodes is smaller than the maximal threshold of candidate groups (i.e., 20). Consequently, when the number of candidate nodes grows from 1 to 20, the network latency increases, similar to the process in classical Raft protocol. However, once the number of candidate nodes reaches the threshold, no more new nodes can join the candidate group, and the network latency will not increase anymore.

\subsection{Sensitivity Analysis}

In this subsection, we conduct a sensitivity analysis to evaluate the effects of changing the {\it c/n} ratio on the performance comparison between our proposed consensus scheme and classical Raft protocols. As stated in the previous section, our protocol significantly reduces the network latency and communication cost, especially when the ratio of {\it c/n} is greater.

\begin{figure}[t]
\centering
\includegraphics[width=0.41\textwidth]{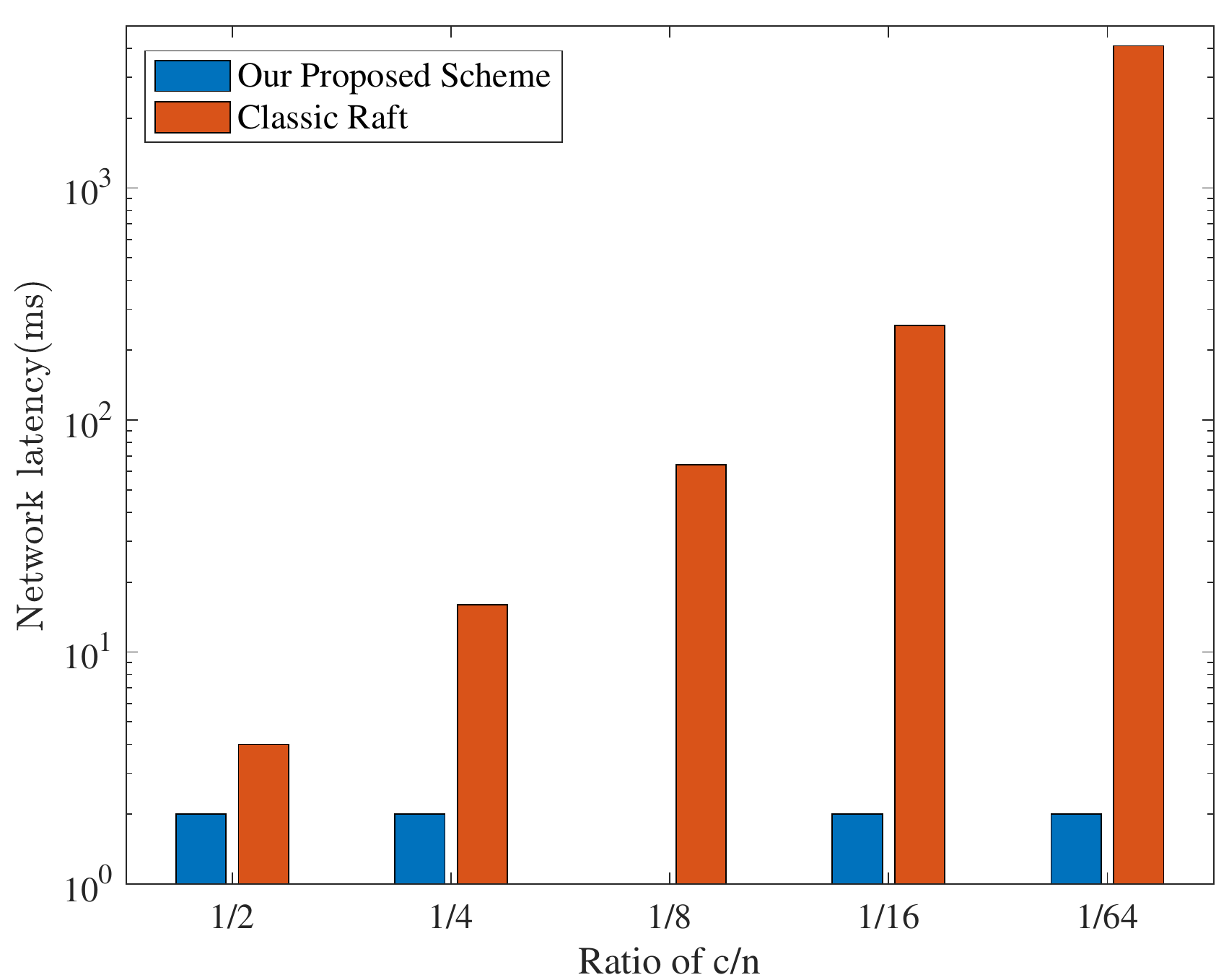}
\caption{Sensitivity analysis with different ratios of candidate nodes and total IoT nodes.}
\label{fig:sensitivity}
\end{figure}

\begin{figure}[t]
\centering
\includegraphics[width=0.41\textwidth]{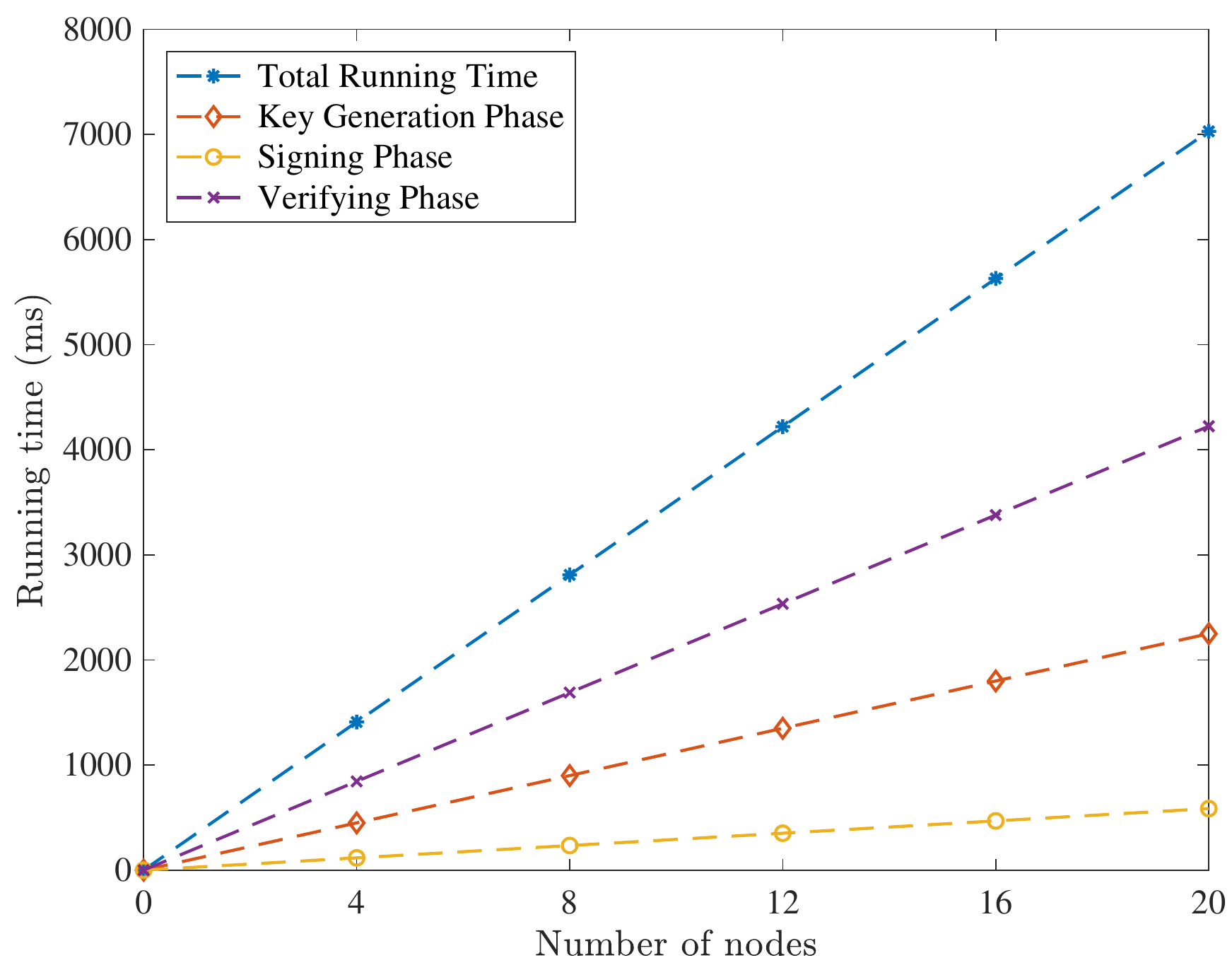}
\caption{Threshold signature scheme running time vs. number of consensus nodes.}
\label{fig:ursa_plot}
\end{figure}

To evaluate the reduction of network latency, we showed multiple groups of experiments with different ratios of {\it Candidate Nodes} to total IoT nodes.  As shown in Fig.~\ref{fig:sensitivity}, our proposed consensus protocol showed significant performance improvements over the classical Raft by increasing the ratio of  {\it c/n}. For instance, when the ratio of {\it c/n} grows from {\it 1/2} to the {\it 1/64}, the network latency of classical Raft protocol will grow exponentially in comparison with our protocol.  
In {\it 1/64} case, the classical Raft protocol's network latency is beyond ${\it 10^3}$ while the new consensus protocol's performance remains at ${\it 10^0 to ~10^1}$ level.

\subsection{Threshold Signature Scheme}

Our threshold signature scheme is developed on the Hyperledger Ursa. 
We first instantiated three consensus nodes to form a lower-layer blockchain network instance. Its unique MAC address can identify each node. Then, we follow Algorithms 1, 2, and 3 in the construction of key generation, signing and verifying processes. In this example, three signatures are signed by the consensus nodes from the lower layer blockchain network and verified by the upper layer blockchain network. For each signature, the average time for key pair generation, signing, and verifying  are 113 ms, 30 ms, and 215 ms, respectively.


We measure the performance of the threshold signature scheme by varying the number of consensus nodes from 4 to 8, 12, 16, and 20 in the lower layer blockchain network. As shown in Fig. \ref{fig:ursa_plot}, the total running time of each phase increases linearly with the increase in the number of consensus nodes in the lower layer blockchain network. Besides, the verifying phase takes additional time than the signing phase because the former requires the computation of pairings based on bilinearlity. In addition, our threshold signature scheme can offer constant running time for signing and verifying phases when varying the length of the identity information.

\section{Conclusion}
Consensus algorithms are the defining mechanism behind the security and performance of blockchain networks. This paper proposed a hierarchical and location-aware consensus protocol for IoT-Blockchain applications.
 Compared to the original Raft protocol, our proposed consensus protocol is scalable by design and reaches consensus faster with lower network overhead and less communication cost. 
 We prototype the experiments by utilizing the Hyperledger Ursa cryptography library to evaluate the threshold signature scheme. The results indicate that the architecture is scalable 
 and suitable for large-scale IoT networks. 
 For future work, we plan to investigate the cross-chain consensus among multiple blockchain systems.

\ifCLASSOPTIONcompsoc
  \section*{Acknowledgments}
  This work is supported by the Fundamental Research Funds for the Central Universities under the Grant G2020KY05109.
\else
  \section*{Acknowledgment}
This work is partially supported by the Fundamental Research Funds for the Central Universities under the Grant G2021KY05101.
  
\fi


\ifCLASSOPTIONcaptionsoff
  \newpage
\fi

\bibliographystyle{IEEEtran}
\bibliography{sig.bib}


\vskip -2\baselineskip plus -1fil 
\begin{IEEEbiography}
[{\includegraphics[width=1.0in,height=1.25in,clip]{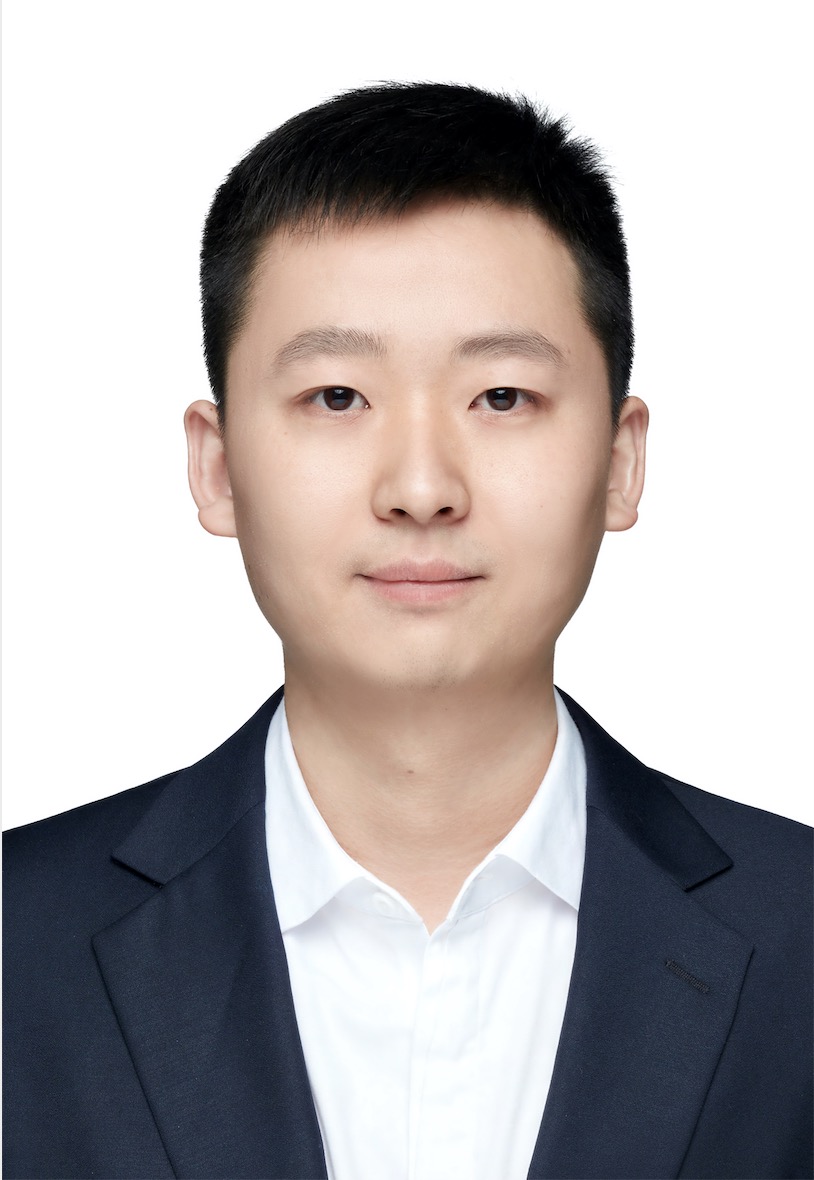}}]{Hao Guo} received the B.S. and M.S. degrees from the Northwest University, Xi'an, China in 2012, and the Illinois Institute of Technology, Chicago, United States in 2014, and his Ph.D. degree from the University of Delaware, Newark, United States in 2020, all in computer science.
He is currently an Assistant Professor
with the School of Software at
the Northwestern Polytechnical University.
His research interests include blockchain and distributed ledger technology, data privacy and security, cybersecurity, cryptography technology, and Internet of Things (IoT). He is a member of both ACM and IEEE.
\end{IEEEbiography}
\vskip -6pt plus -1fil
\begin{IEEEbiography}
[{\includegraphics[width=1in,height=1.25in,clip]{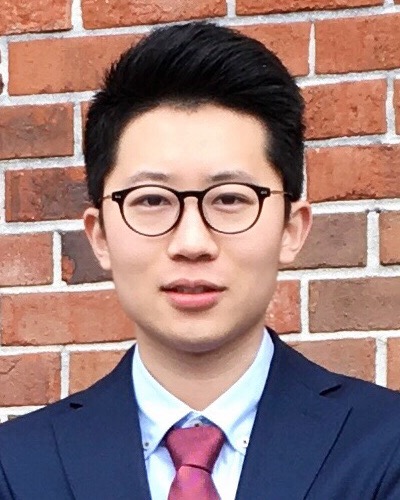}}]{Wanxin Li} received the B.S. and M.S. degrees in computer science from the Chongqing University, Chongqing, China in 2015, and the University of Delaware, United States in 2017, respectively.  He is currently working toward the Ph.D. degree at the University of Delaware. His research interests are in blockchain, intelligent transportation systems (ITS), connected and autonomous vehicles, and Internet of Things (IoT). He is a member of IEEE.
\end{IEEEbiography}
\vskip -6pt plus -1fil
\begin{IEEEbiography}
[{\includegraphics[width=1in,height=1.25in,clip]{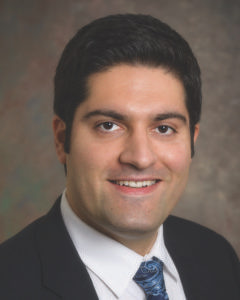}}]{Mark Nejad} is an Assistant Professor at the University of Delaware. His research interests include  network optimization, distributed systems, blockchain, game theory, and automated vehicles. He has published more than forty peer-reviewed papers and received several publication awards including the 2016 best doctoral dissertation award of the Institute of Industrial and Systems Engineers (IISE) and the 2019 CAVS best paper award IEEE VTS. He is a member of the IEEE and INFORMS.
\end{IEEEbiography}
\vskip -6pt plus -1fil







\end{document}